\documentclass[12pt]{article}
\usepackage{a4}
\usepackage{epsfig}
\begin{document}
\title {Thermodynamic properties of the quantum vacuum}
\author{Marco Scandurra \thanks{e-mail: scandurr@df.unipi.it}\\
{\itshape Dipartimento di Fisica}\\
{\itshape Universit\`a di Pisa}\\
{\itshape Via Buonarroti 2, I-56127 Pisa, Italy}}\maketitle

\begin{abstract}
An extension of the fundamental laws of thermodynamics and of  the concept of
entropy to the ground state fluctuations of the quantum fields is studied and
some new results are found. 
At the end a device to extract energy from the vacuum recently proposed by an
author is critically analyzed. 
It is found that no energy can be extracted cyclically from the vacuum.

\end{abstract}
  
\section{Introduction}
The vacuum fluctuations of the fields are one of the most remarkable 
predictions of the relativistic quantum theory of nature. 
The study of the  physical consequences of the quantum vacuum on classical 
bodies began with the discovery of Casimir \cite{Casimir}. 
In the last 50 years the zero point energy has become a challenging subject of
 investigation for scientists of many branches of physics. 
Generalized Casimir forces were studied for all kind of mathematically solvable
geometries, for static and moving boundaries, for hard and penetrable classical
backgrounds. 
Casimir energies were studied in the context of the bag models for hadrons,
 in cosmology and in the background of strong fields 
(gravitational, for instance \cite{Mostepanenko2}. 
The most complete and updated review of the applications of the Casimir effect
is still \cite{Mostepanenko}).
Recently some authors published works dealing with the possibility of using
the Casimir plates as a device to extract energy from the vacuum 
\cite{Pinto,Widom} (see also \cite{cinese}). 
The silence of the ``orthodox'' part of the community expresses the deep 
scepticism on such developments. 
However the policy of ignoring  publications does not contribute to the 
progress of science. 
Discussion is always positive as long as it remains  on the track of scientific
argumentation. 
We also point out that  a rigorous quantum field theoretical analysis of the
ideas lying at the basis of the proposed machines is still lacking.
 This subject gives the opportunity to investigate the  thermodynamics of the 
quantum fluctuations of the vacuum. 
More exactly we would like to address the question: does any thermodynamics in
the vacuum exist? 
The problem was already touched in \cite{Robaschik}, however always in 
connection with finite temperature fields. Furthermore a generalization of the
laws of thermodynamics has been explored in classical and quantum black hole
physics \cite{Bekenstein1,Bekenstein2,Hawking}.
In this paper we are only interested in quantum states with no real particles
and set in flat Minkowski spacetime. 
The question weather the Casimir effect could be used for purposes useful for 
the mankind is an old one and it attracts many adepts to the study of this 
area of physics. 
The author has been frequently asked from students weather the virtual 
particles of the vacuum stand as an infinite reservoir of energy at our 
disposal (a short discussion about this can be found  in section 1.3 of 
\cite{Roberts} and in the references therein). 
With this paper  we intend to cast some light into the topic and remove some 
superstitions. 
For instance it seems necessary to clarify the up to now not deeply 
investigated interplay between Casimir energy and the second law of 
thermodynamics. 
This law is a classical one and a counterpart valid for the sea of vacuum
fluctuations in flat spacetime is by now not known. 

In the next section the possibility of treating the vacuum as a relativistic 
quantum gas with finite thermodynamic variables is discussed. 
In section III the first and the second law of thermodynamics and the concept 
of entropy for the Casimir energy are investigated. 
In section IV a concrete cyclic machine projected to mine the vacuum is 
analyzed. 
In the last section  the results are briefly summarized and discussed. 

\section{The meaning of $\sum_j \frac 12 \omega_j$ in QFT}
        \
Quantum field theory (QFT) predicts the existence of a constant which adds to 
the energy of each system of particles. 
In fact, the complete Hamiltonian of a system reads
\begin{equation}
H=\pm \sum_j \omega_j \left( n_j\ +\  \frac 12 \right) \ , 
\end{equation}
where $\omega_j$ are the one particle energies  and $n_j$ is the number of 
particles in the quantum state $j$ present in the system. 
The sign of the Hamiltonian depends on the statistics, it is positive for 
bosonic fields and negative for fermionic fields. 
The contribution
\begin{equation}
E_0=\pm \frac 12 \sum_j \omega_j,
\end{equation}
which is independent of the occupation number, is the so called vacuum energy 
or zero point energy. 
The momentum operator of the field suffers as well the presence of a divergent
vacuum contribution which is independent of the number of real particles 
present in the system. 
In fact the canonical momentum operator\footnote{we restrict here, for sake 
of brevity, to the scalar case with canonical field operators 
$\hat{\phi}(\vec{x},t)$ and $\hat{\pi}(\vec{x},t)$}
\begin{equation}
\hat{\vec{P}}=-\frac 12 \int d^3x [\hat{\pi}(\vec{x},t)\vec{\nabla}\hat{\phi}
(\vec{x},t)+\vec{\nabla}\hat{\phi}(\vec{x},t)\hat{\pi}(\vec{x},t)]
\end{equation}
reads in terms of the creation and annihilation operators 
$\hat{a}^\dag_{\vec{p}} $ and $\hat{a}_{\vec{p}} $ in the Fock space
\begin{equation}
\hat{\vec{P}}=\frac 12 \int d^3p\ {\vec{p}} (\hat{a}^\dag_{\vec{p}}\hat{a}_{\vec{p}} +  \hat{a}_{\vec{p}} \hat{a}^\dag_{\vec{p}}),
\end{equation}
which in the discrete formulation, after applying the usual commutation 
relations, becomes
\begin{equation}
\hat{\vec{P}}=\sum_j \vec{p}_j(\hat{n}_j+ \frac 12)\ ;
\end{equation}
since the vacuum momenta are isotropically distributed in all directions of
empty space, the total momentum results to be zero. 
However the total vacuum pressure does not vanish so trivially. 
Contribution (2)  has undergone a number of different interpretations: 
for many it represents a mere mathematical protuberance to the Hamiltonian due
 to the imperfect formulation of the theory and with no physical meaning. 
For others the second addendum in the parenthesis in (1) has the same dignity 
as the first addendum and  it must have precise physical consequences. 
The discovery of the Casimir effect and its recent experimental verifications 
\cite{Lamoreaux,Mohideen} seems to let no doubts nowadays on the reality of the
{\itshape renormalized} vacuum energy, 
i.e. small differences in the zero point energy (2) are indeed physically real
quantities. 
However some mystery about the nature of the whole contribution  
$\frac 12 \sum_j \omega_j$ remains. 
If this contribution has to be considered as  unphysical on the other hand it 
can not be neglected in the calculations. 
Furthermore it is  unclear how the difference between two unphysical quantities
\footnote{for instance the vacuum energies $\frac 12 \sum_j \omega_j^{IN}$ and 
$\frac 12 \sum_j \omega_j^{OUT}$ inside and outside a Casimir cavity} could be
a physical one!
The trick is hidden in the well known procedure of renormalization. 
Still the subtraction of the poles of a function  possesses a mathematical 
and not a genuine physical motivation.
If we examine equation (1) we see that the addenda between the parenthesis are 
dimensionless numbers indicating the number of quanta of the field. 
However, while the first addendum depends on the quantum index $j$, the second
 addendum does not. 
So if we have three particles in the state $j=1$ each with energy $\omega_1$ 
formula (1) tells us we have also a {\itshape half} particle in that state with
energy $\omega_1$. 
In each eigenstate of the Hamiltonian we have a half particle  which is present
even if the number of ``integer'' particles in the Fock space is zero. 
This gas of ``half particles'' is the quantum vacuum. 
Unlike the gas of Fock particles which obeys the statistics 
(Bose-Einstein or Fermi-Dirac) of the field at finite temperature, 
the gas of half particles does not have a frequency and temperature dependent 
distribution function. 
The distribution function is merely $\frac 12$.
In the Fock state $|0>$ only this entity, or virtual gas, fills the Minkowski 
space. 
In order to give a meaning to this statement let us investigate whether in each
finite volume of empty space there exist a finite vacuum energy.
For simplicity let us take a slab of empty space of size $L$x$L$x$d$, 
with $L>>d$ in the vacuum of the electromagnetic field. 
We impose periodic boundary conditions on the surface of the slab. 
No material boundary delimits this volume, what we are assuming here is that 
the electromagnetic waves propagate continuously all through space. 
The total energy present in the volume $V=L$x$L$x$d$ is given by
\footnote{the reader is assumed to be familiar with this type of expressions}
\begin{equation}
E^{slab}_0\ =\ \sum_n \frac 12 \omega_n^{slab}\ =\ \frac{\hbar c L^2}{4 \pi^2}\int_0 ^\infty dk_1 dk_2 \sum_n^\infty \sqrt{k_1^2 +k_2^2+(2 \pi n/d)^2}\ . 
\end{equation}
We regularize this expression with the help of zeta-functional techniques \cite{ZetaRegularization}, i.e. we introduce a complex power $s$
\begin{equation}
E^{slab}_0\ =\ \frac{\hbar c L^2}{4 \pi^2}\int_0 ^\infty dk_1 dk_2\sum_n^\infty \sqrt{k_1^2 +k_2^2+(2 \pi n/d)^2}^{(1-2s)} \ .
\end{equation}
For $\Re s>3/2$ this expression converges. 
The analytic continuation to $s=0$ can be performed as  follows. 
We introduce polar coordinates $(k, \varphi)$ in the $(k_1, k_2)$-plane. 
Because of the presence of the regularization parameter, the integration over 
$k$ and $ \varphi$ can  be performed. 
With the use of the Riemann zeta function $\zeta_R=\sum_{n=1}^\infty n^{-s}$ 
we obtain
\begin{equation}
E^{slab}_0=\frac{\hbar c L^2}{2 \pi}\frac{(2\pi/d)^{3-2s}}{2s-3}\zeta_R(2s-3).
\end{equation}
For $s=0$, with $\zeta_R(-3)=-B_4/4=1/120$, the energy turns out to be
\begin{equation}
E^{slab}_0=-\frac{\hbar c L^2}{d^3}\frac{\pi^2}{360}\ ,
\end{equation} 
which it is just half of the Casimir energy between to perfectly conducting 
plates.
This means that to each finite volume of empty Minkowski space it can be 
associated a finite vacuum energy density. 
This quantity exists even if no material object is put into the vacuum to 
perturb it. 
The derivative of (9) with respect to $d$ yields a pressure
\begin{equation}
P^{slab}_0=-\frac{\hbar c }{d^4}\frac{\pi^2}{120}\ ,
\end{equation} 
which is a quantity intrinsic of the unbounded volume $V$ as well. 
We find then ourselves in the situation of having two non-vanishing fundamental
thermodynamic variables associated to each finite region of empty space.
 However these variables depend on the parameter $d$, which makes them 
ambiguous. 
Furthermore energy density and pressure change with the geometry of the 
considered volume. 
For this reason no general equation of state binding $V$, $E_0$ and $P_0$ can 
be written.

\section{The fundamental laws of thermodynamics}
We try here to extend  the  principles and the equations of the classical 
theory of heat to the renormalized energy of the fluctuating quantum fields. 
We  start with the second law of thermodynamics. In the formulation of Thomson 
(Lord Kelvin \cite{Kelvin}), 
the second law of thermodynamics prevents from extracting energy from a single 
bath of heat at a certain temperature. 
This would not violate conservation of energy, however, it would decrease the 
total entropy of the system. 
Such a transformation is, to our knowledge, forbidden by nature
\footnote{more exactly, a decrease in entropy characterizes extremely 
improbable transformations.}. 
To extract energy we need at least two baths of heat which are at two 
different temperatures. 
Then, energy can be taken from the hotter bath releasing a part of it to the 
bath at lower temperature. 
No matter how sophisticated the mining machine is, the ratio of the net 
gained energy over the total extracted energy cannot be larger than the 
Carnot coefficient $\eta<1$, given by
\begin{equation}
\eta=\frac{T_1-T_2}{T_1},
\end{equation}
where $T_1$ and $T_2$ are the temperatures of the hot and of the cold bath 
respectively. The coefficient $\eta$ expresses the efficiency of a thermal 
machine \cite{Carnot}. 
It also fixes mathematically the statement of the second law of thermodynamics
\footnote{A rigorous mathematical formulation of the second law of 
thermodynamics is the following: 
for each transformation which involves a heat variation $\delta Q$ in a bath 
at temperature $T$, the quantity $\delta Q/T=dS$ is an exact differential 
(see for instance \cite{Pauli}).}. 
We would like to investigate if an analogous of formula (11) exists for the 
zero point fluctuations.
It happens that to energy (9), as well as to the Casimir energy inside two 
perfectly conducting plates, it can be associated a finite effective 
temperature. 
In fact the energy density $-\frac{\pi^2 \hbar c}{d^4}\frac{1}{360}$ can be 
given by the integral
\begin{equation}
-\frac{\pi^2 \hbar c}{d^4}\frac{1}{360}\ =\ \frac{\hbar}{\pi^2c^3}\int_0^\infty \frac{d\omega\ \omega^3}{e^{\frac{\hbar\omega}{k_BT_{eff}}}-1}\ ,
\end{equation}
where $k_B$ is the Boltzmann constant and  $T_{eff}\sim d^{-1}$ is the so 
called {\itshape effective temperature} of the vacuum. 
The r.h.s of (12) is an integration over the frequencies of a thermal bath of 
photons having temperature $T_{eff}$. 
This means that under certain circumstances the renormalized vacuum formally 
coincides with a photon gas having a Planckian spectrum and a finite 
temperature
\footnote{This property was first analyzed in \cite{Ford-Spec} for the one 
dimensional cavity. 
It is also widely treated in chapter 5 of reference \cite{Mostepanenko}. 
The main critic that can be raised to this idea is that the Casimir energy 
inside a cavity is not homogeneously distributed like it is the black body 
radiation. 
It has been shown (see again \cite{Mostepanenko}) that the spatial 
distribution of the Casimir energy density has asymptotes on the surface of 
the boundaries}. 
The exact value of  $T_{eff}$ for the periodic slab reads\footnote{There are 
actually four solutions to eq. (12). 
The solutions differ in the coefficients and in the sign, keeping the same 
dependence on $d$. We have taken here one of the two positive solutions. 
The result which follows in a few lines is independent of the solution chosen}
\begin{equation}
T_{eff}^{periodic}\ =\ \frac{(-1)^{3/4}}{2^{3/4}3^{1/4}}\ \frac{c\hbar}{k_B d}
\end{equation}
and for the volume inside the conducting plates
\begin{equation}
T_{eff}^{conduct}\ =\ \frac{(-1)^{3/4}}{2^{3/4}3^{1/4}}\ \frac{c\hbar}{2 k_B d}\ \ .
\end{equation}
They are both imaginary. However, if one only considers the moduli of the 
energies on both sides of (12), one finds the real temperatures
\begin{equation}
T_{eff}^{periodic}\ =\ \frac{1}{2^{3/4}3^{1/4}}\ \frac{c\hbar}{k_B d}
\end{equation}
\begin{equation}
T_{eff}^{conduct}\ =\ \frac{1}{2^{3/4}3^{1/4}}\ \frac{c\hbar}{2 k_B d}\ \ .
\end{equation}
Now we want to find the Carnot coefficient $\eta_0$ for a machine working 
between two Casimir cavities with sizes $d_1$ and $d_2$, with $d_1<d_2$. 
We will adopt formula (11) understanding $T_1$ and $T_2$ as effective 
temperatures. 
Let us first consider the case of the perfectly conducting plates and let us 
take up the real temperatures given by (16) for $d=d_1$ and $d=d_2$. 
Inserting them in to (11) we find the  simple result
\begin{equation}
\eta_0\ =\frac{d_2-d_1}{d_2},
\end{equation}
using the imaginary temperatures given by (14) the factors $i$ simplify and we 
obtain the same result. 
Furthermore, we calculated $\eta_0$ for the case of periodic boundary 
conditions and we obtained again result (17). 
The coefficient (17) has to some extent a universal meaning, it does not 
depend on the boundary conditions, it only depends on the parameter $d$. 
We would also like to note that an analogous coefficient must exist and  could 
be calculated in black hole physics, where an effective temperature is 
associated to the mass of an evaporating singularity.
We would like to make a further remark: 
Eq.(11) does not depend on the kind of statistics of the considered field. 
For massless Bose and Fermi fields the energy density is $u\sim T^4$ and  
equation (11) can be rewritten as follows:
\begin{equation}
\eta\ =\ 1-\left( \frac{u_1}{u_2}\right)^{1/4}, 
\end{equation}
where $u_1$ is the energy density of the hotter thermal bath and $u_2$ the 
energy density of the cooler one. 
By means of (18) the Carnot efficiency in vacuum can be calculated without the 
use of the effective temperature, by simply inserting the Casimir energy 
density of the chosen configuration. 
Let us find the coefficient in the case of unpenetrable spherical shells. 
The renormalized Casimir energy for a perfectly conducting hollow sphere with 
radius $R$ is the well known \cite{Boyer}
\begin{equation}
E_0^{shell}\ =\ +0.09235\ \frac{h c}{2R}\ .
\end{equation} 
The corresponding energy density reads
\begin{equation}
u^{shell}\ =\ +0.09235\ \frac{3 hc}{8\pi R^4}\ .
\end{equation}
Inserting in (18) the energy density (20) taken for two shells of radii $R_1$ 
and $R_2$ with $R_1<R_2$, we find
\begin{equation}
\eta^{shell}=1-\frac{R_1}{R_2}\ ,
\end{equation}
which is again result (17). 
The coefficient $\eta_0$ seems then to be independent on the geometry of the 
system. 
Equation (17) establishes a connection between quantum fluctuations of the 
fields and the second law of thermodynamics, 
giving the efficiency of a Carnot machine in vacuum. 
The coefficient $\eta_0$ tell us that if $d_2$ (i.e. the size of the ``cold'' 
cavity) reaches infinity, a hundred per cent of the available energy can be 
transformed in to work for human purposes. 
However, this means  100 per cent of the renormalized energy $u_1$ confined in 
the small cavity  and not of the huge energy 
$\frac 12 \sum_j^\infty \omega_j$ outside it.
Now some authors suggested that a machine working between two reservoirs at 
different Casimir energy densities should be able to extract an unlimited 
quantity of energy from the vacuum. 
In the next section we will analyze this possibility in detail.

An important consequence of the second law of thermodynamics is that, for 
every transformation,  the variation in entropy $S$ is always positive or zero.
This drives the whole universe toward a state of maximum entropy determining 
the time direction of the events. 
As Clausius pointed out \cite{Clausius}, the universe marches toward its 
thermal death: 
a global state in which the total entropy has reached its maximum. 
In this universe  no thermodynamic  transformation can take place. 
In the definition given by  Clausius, the variation in entropy for an 
isothermal  transformation is calculated by:
\begin{equation}
dS\ =\ \frac{dQ}{T}\ .
\end{equation}
Boltzmann \cite{Boltzmann} gives a more elegant definition, involving purely 
probabilistic considerations for the entropy of a system which is in the state 
$s$:
\begin{equation}
S\ =\ k_BN\ln W_s\ .
\end{equation}
Here $N$ is the number of particles of the system and $W_s$ is the probability 
for the state $s$. It can be shown that definition (22) and (23) are
equivalent \footnote{In statistical quantum mechanics, the entropy is commonly
expressed as $S=k\ln Z +kT\ \frac{\partial \ln Z }{\partial T}$, where $Z$ is 
the sum-over-states: $Z=\sum_n \exp[-E_n/kT]$, the quantities $E_n$ being the 
energy eigenvalues for all the different steady state solutions $n$ of which 
the system is capable. For a proof of the validity of equations (11), (22) and 
of eq.(24) in quantum statistical mechanics, see sections 129, 130 and 131 of \cite{Tolman}}.
The author of the present paper finds of interest to investigate the variation 
in entropy for transformations in vacuum involving Casimir cavities.
We find ourself immediately in the difficulty of defining the quantity $dQ$. 
In fact we do not have any heat exchange in Casimir like situations. 
It is a very remarkable property that the walls of a Casimir cavity do not 
transmit energy between the two regions of space which are at two different 
energy densities.
Given this peculiarity, it seems necessary to redefine the fundamental equation
of the first principle of thermodynamics, which in classical physics reads:
\begin{equation}
dQ=dU+dW.
\end{equation}
This states that in each system a variation of heat $dQ$  is partially 
transformed into work $dW$, while a part of it is used to change the internal 
energy of the system $dU$.
When dealing with the quantum vacuum, the internal energy can always be 
defined by means of the renormalized (Casimir) energy and eventually by means 
of the effective temperature. 
A work $W$ can always be defined when the size of the cavity is changed, 
therefore, we can always define ``adiabatic'' transformations, as 
transformations for which $dQ=0$ and $-dU=dW$. 
We now ask ourselves  whether an energy variation can take place when the size 
of the cavity remains fixed. This is indeed the case when the boundary 
conditions are changed. 
A change in the physical properties of the walls of the cavity results in a 
change of the Casimir (internal) energy  and, if the cavity is free to expand 
or contract this change will also produce work. 
It corresponds to a heat exchange $dQ$ in classical physics. 
The first law of thermodynamics may then be rewritten as:
\begin{equation}
dB=dC+dW\ ,
\end{equation}
where $dB$ is the energy expended to the vary the boundary conditions and $dC$ 
is the variation in the renormalized vacuum energy inside the cavity. 
If we vary the boundary conditions of a cavity while keeping its size 
unchanged, all the energy $dB$ is turned into Casimir energy and no work is 
performed. 
We have then an isochoric transformation:
\begin{equation}
dB=dC\ .
\end{equation}
If otherwise we compensate the variation of the boundary conditions with a 
change in the volume of the cavity, so that the Casimir energy remains 
unchanged, we have an ``isothermal'' transformation:
\begin{equation}
dB=dW\ .
\end{equation}

In full analogy with the classical theory, the law of entropy may be rewritten 
as follows:
\begin{equation}
dS_c = \frac{dB}{T_{eff}}\ ,
\end{equation}
where  $ S_c $ indicates a kind of modified or Casimir entropy. We want to 
study the sign of $dS_c $. We will not accomplish rigorous calculations here,
 we only propose an  approach to the study of $S_c$. Let us first consider  
the spherical 
geometry. 
For dimensional reasons, a conducting spherical shell with radius $R$, in the 
electromagnetic vacuum, has an effective temperature of the form
\begin{equation}
T^{spher}_{eff}\ \sim \ \sigma \frac{hc}{k_B R}\ , 
\end{equation}
where $\sigma$ is a number. 
Let us suppose to have a spherical shell with radius $R_1$ and with a boundary 
condition (a hardness) ${\cal B}_1$ on its surface which is close to the 
perfect reflectivity. 
Inside the sphere we have a finite and positive Casimir energy to whom it can 
be associated an effective temperature of the form (29)
\footnote{the calculation of this temperature is of course a hard  task and a 
simple analytical form in terms of $R_1$ and ${\cal B}_1$ may even not exist}. 
We now let the shell expand under the effect of the repulsive Casimir force 
until it reaches the radius $R_2$. 
During this process a work is done. 
At a same time we change the boundary conditions toward a higher hardness 
${\cal B}_2$, so that the two effects compensate each other and the 
renormalized energy and effective temperature remain unchanged. 
In passing form the boundary condition  ${\cal B}_1$ to the boundary condition 
${\cal B}_2$ an energy $\Delta B$ must be expended, 
in fact more frequencies are trapped inside the shell and the Casimir 
mechanism, generating positive energy is enhanced. The change in entropy is 
given by
\begin{equation}
\Delta S_c^{spher}=\int_{B_1} ^{B_2} \frac{dB}{T_{eff}^{spher}}\ .
\end{equation}
Being $B_2>B_1$ and $T_{eff}^{spher}>0$, this quantity is clearly positive.\\
In the case of two parallel plates embedded in the vacuum, we calculate the 
entropy variation for the following transformation: 
We let two  highly reflective boundaries initially at a distance $d_1$ move 
one toward another under the effect of  the attractive Casimir force and we 
compensate the change in Casimir energy by a reduction of the hardness of the 
boundaries (${\cal B}_2 \rightarrow {\cal B}_1$). 
The effective temperature remains unchanged and we have an entropy variation
\begin{equation}
\Delta S_c^{plat}=\int_{B_2} ^{B_1} \frac{dB}{T_{eff}^{plat}}\ .
\end{equation}
This quantity is positive, since the work we must do to diminish the 
reflectivity of the plates is positive.
If we let the parallel plates or the spherical shell freely move under the 
effect of the vacuum forces, without any change in the boundary conditions, 
we have an adiabatic transformation ($dB=0$) and the entropy variation is $0$. 

We have then introduced in (28) a sort of entropy, which applies to quantum 
problems where the boundary conditions play a relevant role. 
We found (at least in the above considered class of examples) that in vacuum  
this quantity never decreases.

\section{Casimir machines}
The principal idea in an energy  extraction cycle which adopts the Casimir 
effect is to let the  Casimir force do a work and take the energy  released in 
this process. 
For instance we can let the Casimir plates freely ``fall'' one toward another 
from a distance $d_1$ to a distance $d_2$. 
The plates may ``impact''  at the distance $d_2$, where the repulsive 
intermolecular forces become overwhelming and  the kinetic energy could be 
transformed into electrostatic energy to be stored in a capacitor. 
The stored energy is very small. 
To restart the process again one must bring the two plates apart. 
This costs  an amount of work which is $C(d_1)-C(d_2)$, where $C(x)$ represents
the Casimir energy of two perfectly conducting plates kept at a distance $x$. 
This amount of work exactly compensates the gained energy. 
We will now being assuming that this compensation is not due to any fundamental
law of nature, but rather  to a lack of our extraction device. 
We shall therefore consider a more  complex extraction cycle, like, for 
instance,  that proposed in \cite{Pinto}. 
In that paper, ``adiabatic'' and ``non adiabatic'' transformations are used to 
form a closed cycle which looks very similar to a Carnot cycle when displayed 
in a diagram having the plates separation on the abscissa and the Casimir 
pressure on the ordinate. 
The machine proposed in \cite{Pinto} uses the electromagnetic vacuum and two 
parallel plates whose dielectric (reflectivity) properties vary periodically, 
enhancing or diminishing the Casimir force.  
The change in the dielectric properties of the medium is accomplished by 
illuminating the medium with  suitable radiation. 
The flow of radiation through the boundaries made of semi conducting material 
ionizes the atoms of the medium, causing an increase in the number of free 
electrons and in the conductivity properties of the boundaries. 
We are not interested here in the details of optics, therefore this description
is sufficient for our purposes. 
We will analyze a simplified version of the machine proposed in \cite{Pinto}. 
It is displayed in Fig. 2. 
At phase A two highly transparent thin boundaries are kept at a large distance 
from each other, so that the Casimir force/energy is negligible. 
At a certain time (picture B) two suitable lamps are turned on before the two 
boundaries. This increases the conductivity of the medium and the modes of the 
vacuum will be scattered by the boundaries. 
Previously the boundaries were almost totally penetrable and no scattering took
place. 
The scattering of the modes inside the cavity causes the creation of stationary
waves and the change of the energy spectrum which activates the Casimir 
mechanism; this represents a non-adiabatic transformation.  
The plates will attract and the vacuum will be doing work moving the two 
boundaries and bringing them closer, this  motion  represents an adiabatic 
transformation. 
In the phase C the motion is stopped at an arbitrary  distance $d_2$. 
In order to drastically reduce the Casimir attraction, the lamps are turned 
off (picture D). 
The boundaries become again almost fully transparent and they can be taken away
from each other with a negligible amount of external work. 
When the plates are sufficiently distant  the lamps are turned on again and a 
new cycle may start. 
This is an intriguing apparatus in which the reduction of the Casimir potential
by turning off the lamps at the stage D of the cycle seems to happen at 
absolutely no cost of energy. 
During the whole cycle, the only energy that  must be yielded to the system 
from outside is that necessary to ionize the atoms of the medium in order to 
increase the reflectivity of the boundaries. 
However, when the lamps are turned on at the beginning of the cycle (stage B),
 the Casimir potential is reduced since we create a negative energy density 
inside the plates, this should not cost us energy, on the contrary energy 
should be released, since the system tends toward the state of lowest 
potential.  
Now,  is hard to imagine how, by emitting photons, we can collect energy from 
the vacuum! 
On the other hand when turning off the lamps (stage D) some external work is 
done, because the Casimir potential is increased (remember that the sign of 
the potential is negative), however one could hardly imagine to yield energy 
to the environment by stopping emitting photons in it!  
To get rid of this paradox let us analyze the entire cycle again, this time 
neglecting all forces and energy changes which have nothing to do with the 
Casimir  effect and which certainly obey conservation laws. 
At stage A,  when the transparent boundaries are kept at a large distance, a 
weak Casimir force will tend to bring the potential to the lowest state, i.e. 
to increase the modulus of the Casimir energy. 
This includes a weak force which will tend to ionize the atoms of the medium in
order to render the boundary more reflective. No lamp is turned on. 
As the boundaries begin to move one toward another, the Casimir force gets 
larger and a larger number of atoms is ionized. 
The decrease in potential energy is compensated by an increase in kinetic 
energy of the plates. 
As the boundaries reach the minimum distance and the kinetic energy is stored 
for human purposes, the potential energy reaches its minimum, the Casimir 
force reaches its maximum and a large part of the atoms of the boundaries is 
kept ionized. 
Now if we wish to restore the initial conditions and restart the cycle again 
we must first render the material of the boundaries transparent, to do so we 
have to perform  work against the forces of the vacuum which tend to keep the 
electrons free and the material highly reflective. 
This is the crucial point. 
The work we must do to recombine the electrons with the atoms is of course 
larger when the distance of the plates is small and it is presumably equal to 
the total energy released by the vacuum during the  compression phase. 
One may object that the vacuum  releases energy in ionizing the atoms  just 
like it releases kinetic energy in bringing the plates closer. 
This is indeed true. 
We could markedly distinguish between 1. the energy released from the vacuum 
in bringing two dielectrics closer and 2. the energy released from the vacuum 
to render two close dielectric boundaries highly reflective. 
However the humans who want to restore the initial conditions will have to do 
the opposite jobs as well: 
they will have to 1. recombine the free electrons against the forces of vacuum,
2. take the two weakly reflective boundaries away one from another against the
forces of vacuum. 
The two contributions may smoothly overlap during the transformation, but the 
final result is the same: the energy budget is zero at the end of the cycle. 
The introduction of other variables in the system, like the electrostatic 
forces of the atoms  which tend to keep the electrons bounded, or the photons 
of the lamp which  help in the ionization, does not alter this conclusion:  
no energy is extracted cyclically by means of a machine mixing ``adiabatic'' 
and ``non-adiabatic'' Casimir transformations.     

\section{Conclusions}

Investigating the thermodynamic properties of the ground state fluctuations of 
the quantum fields a number of interesting features appear. 
We have studied the possibility of  associating finite thermodynamic quantities
like energy density and pressure, to a finite region of empty space; this is 
possible by renormalization, but the thermodynamic variables depend on the 
geometry of the chosen region. 
We have also discussed the first and the second principle of thermodynamics in
connection with Casimir energies. 
We found an expression for  the thermal efficiency of a machine working 
between two cavities confining the electromagnetic vacuum at different energy 
densities. 
We also faced the problem of entropy in vacuum physics. 
We defined a modified version of the entropy $S_c$, which depends on the 
boundary conditions introduced in a system. 
We reached a result which suggests that for plane and spherical geometries the
variation in $S_c$ is always positive. 
We note that this result is not in contradiction with Nernst theorem 
\footnote{The problem of the validity of the Planck-Nernst theorem is also 
discussed within black holes thermodynamics \cite{Wald}}. 
The latter in fact applies to the classical concept of entropy. 
The quantity $S_c$ not only takes into account boundary condition effects, but 
it picks up the effective temperatures of the bounded regions instead of the 
value T=0 which classical physics trivially  attributes to the vacuum. 
The effective temperature may be then regarded to as a kind of 
{\itshape renormalized temperature}, which allows to construct a thermodynamic 
theory of the vacuum fluctuations \footnote{We have here another interesting 
analogy with the physics of black holes: In classical gravitation the 
temperature of a black hole is $T=0$ and no energy escapes the event horizon. 
However, when quantum effects are taken into account a finite temperature must
be assigned to the black hole, which radiates particles thermally.}.
    
In the last section we have  critically examined the possibility of extracting 
energy from the zero point fluctuations, reviewing the device proposed in 
\cite{Pinto} and showing that a closed cycle with a net positive energy budget 
cannot be realized. 
A Casimir machine transforms into work only a small and finite quantity of 
energy, namely the Casimir energy involved with the geometry the cavity. 
Once this quantity is  consumed, there is no way to restore the 
initial conditions of the cavity by varying some physical parameters without 
spending out all the gained energy. 

The hypothesis that the vacuum contribution  $\frac 12 \sum_j^\infty \omega_j$,
predicted by quantum field theory, is an infinite heat reservoir from which 
we can take unlimited energy for our purposes does not find confirmation in 
the present paper.  
We are only able to manipulate renormalized energies, which the vacuum makes 
us available. 
It may be nonetheless possible to rescue the concept of the renormalized vacuum
as a kind of quantum relativistic gas pervading empty space, 
in this case the principles of thermodynamics seem to be still valid, when some
opportune modifications are included. 
Accepting this picture many questions still remain open,  the author hopes to 
return on them in a future paper.

\section{Acknowledgments}
I would like to thank the Italian foundation 
ONAOSI for support during this research.

\vfill\eject

\begin{figure}[ht]\unitlength1cm
\begin{picture}(10,10)
\put(0,-5){\epsfig{file=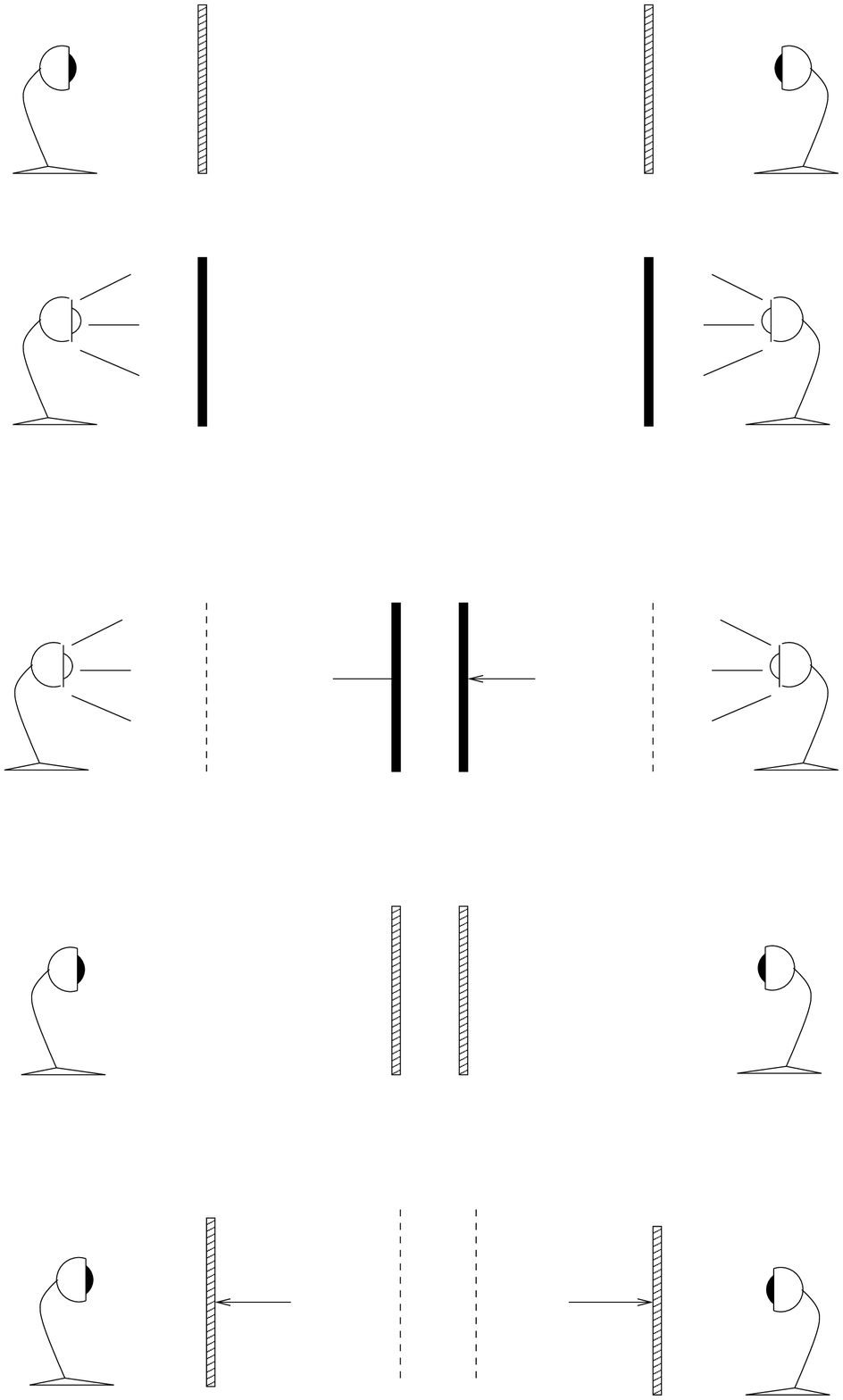,width=14cm,height=17cm}}
\put(-2,10.7){ A }
\put(-2,7.7){ B }
\put(3.9,7.7){\small highly reflective}
\put(3.9,10.8){\small highly transparent }
\put(0.0,11.66){\footnotesize ionizing lamp }
\put(-2,3.4){ C }
\put(-2,-0.3){ D }
\put(-2,-4.1){ E }
\end{picture}

\vspace{7cm}
\caption{A cycle for the extraction of energy from the vacuum by using  
``non adiabatic'' transformations.} 
\end{figure}

\end{document}